\documentclass[
    ,final            
  ]
  {aipproc}

\layoutstyle{8x11single}

\def\be{\begin{equation}}
\def\ee{\end{equation}}
\def\bea{\begin{eqnarray}}
\def\eea{\end{eqnarray}}

\begin{document}

\title{Strongly magnetized iron white dwarfs and the total lepton number violation}


\author{V.B.~Belyaev}{
  address={Bogolyubov Laboratory of Theoretical Physics, Joint
Institute for Nuclear Research, Dubna  141980, Russia} }
\author{P. Ricci}{
  address={Istituto Nazionale di Fisica Nucleare, Sezione di
Firenze, I-50019 Sesto Fiorentino (Firenze), Italy } }
\author{F.~\v{S}imkovic}{
  address={Department of Nuclear Physics and Biophysics,
Comenius University, Mlynsk\'a dolina F1, SK--842 15, Bratislava,
Slovakia}, altaddress={Bogolyubov Laboratory of Theoretical Physics,
Joint Institute for Nuclear Research, Dubna  141980, Russia} }
\author{J.~Adam,~Jr.}{address={Institute of Nuclear Physics ASCR, CZ--250 68 \v{R}e\v{z}, Czechia}}
\author{M.~Tater}{address={Institute of Nuclear Physics ASCR, CZ--250 68 \v{R}e\v{z}, Czechia}}
\author{E.~Truhl\'{\i}k}{address={Institute of Nuclear Physics ASCR, CZ--250 68 \v{R}e\v{z}, Czechia}}

\begin{abstract}
The influence of a neutrinoless electron to positron conversion on a
cooling of strongly magnetized iron white dwarfs is studied.
\end{abstract}

\maketitle


\section{Introduction}

Continuously improving description of the cooling of white dwarfs
(WDs) and precise measurements of their luminosity curve open a
possibility of their use as an elementary particle physics
laboratory. Thus, Isern {\it et al.} \cite{IEA} suggested to test
possible existence of axions by studying the WDs luminosity
function. Analogously, we analyze the influence of lepton number
violation on the luminosity of strongly magnetized iron WDs
(SMIWDs). The existence of Majorana type neutrino would imply the
lepton-number-violating process of electron capture by a nucleus
$X(A,Z)$:
\begin{equation}
e^- + X(A,Z) \rightarrow X(A,Z-2) + e^+\,,  \label{EMEPC}
\end{equation}
which is an analogue of the netrinoless double beta-decay,
intensively studied  these days \cite{VES}.

The description of the strongly magnetized white dwarfs (SMWDs: we
use this acronym for the WDs with the magnetic field in the core
higher than a critical one ${\cal B}_c$=4.414 $\times 10^{13}$ G) is
based on a strongly magnetized cold degenerate electron gas
\cite{KM}. The theory follows from the Landau quantization of a
motion of electrons in  homogenous magnetic field, usually taken to
point along the z-axis \cite{LL}, with a modification to the case of
very strong ones \cite{LS}. In systems with small number of Landau
levels, which is restricted by the strength of the magnetic field
and by the Fermi energy E$_{\,\mathrm{F}}$ of the electron gas, the
mass of the SMWD can be in the range (2.3 - 2.6)\,M$_\odot$, where
M$_\odot$ is the solar mass. That is, the strong magnetic field can
enhance the energy of the electron gas to such a level that its
pressure allows the SMWD to have a mass larger than the
Chandrasekhar-Landau (ChL) limit of 1.44\,M$_\odot$ \cite{C}.

We apply the theory of SMWDs to estimate, whether the influence of
the double charge exchange reaction (\ref{EMEPC}) on cooling of the
SMIWDs can be detectable. The threshold for this reaction with the
initial nucleus being $^{56}_{26}$Fe and the final one
$^{56}_{24}$Cr is $\Delta$=6.33 MeV, thus in the SMIWDs with
E$_{\,\mathrm{F}}\,\ge\,\Delta$ this reaction can take place. We
consider
$\epsilon_{\,\mathrm{F}}$=E$_{\,\mathrm{F}}$/m$_e$\,c$^2$=20,\,46,\,90,\,150
and 200 and choose the strength of the magnetic field so that the
value of $\gamma={\cal B}/{\cal B}_c$ allows only for the ground
Landau level.

\section{Theory of strongly magnetized white dwarfs}

Theory of the SMWDs is  discussed in detail in Refs.\,\cite{KM}. In
the relativistic case, one solves the Dirac equation, obtaining for
the electron energy the equation:
\be E_\nu=m_e c^2 \left[1 + \left(\frac{p_z}{m_e c}\right)^2 +
2\nu\gamma \right]^{1/2} \, . \label{ENU} \ee
Here $m_e $ is the electron mass, $c$ is a velocity of light, $p_z$
is the electron momentum along the z-axis, $\nu=l+1/2+\sigma$ labels
the Landau levels with the principal number $l$, $\sigma=\pm 1/2$.
The ground level ($\nu$=0) is obtained for $l$=0 and $\sigma$=-1/2,
and it has the degeneracy factor g$_\nu$=1. Other Landau levels
posses the degeneracy factor g$_\nu$=2. While the  density of
electron states in the absence of the strong magnetic field is given
as $2/(2\pi\hbar)^3\,d^3p$, the presence of such magnetic field
modifies the number of electron states for a given level $\nu$ to $2
g_\nu e{\cal B}/[(2\pi)\hbar)^2 c]dp_z$. Then the sum over the
electron states in the presence of the strong magnetic field is
given by:
$$
\sum_E \rightarrow \sum_\nu \frac{2 e{\cal B}}{(2\pi\hbar)^2 c}\,
g_\nu \int\,dp_z = \frac{2\gamma}{(2\pi)^2\lambda^3_e}\, \sum_\nu
g_\nu \int d \frac{p_z}{m_e c} \ ,
$$
where $\lambda_e=\hbar/m_e c$ is the electron Compton wavelength.

In what follows, we consider the case of the ground Landau level,
for which the Fermi energy $\epsilon_{\,\mathrm{F}}$ and the Fermi
momentum $x_{\,\mathrm{F}}=p_{\,\mathrm{F}}/m_e c$ are obtained
directly from Eq.\,(\ref{ENU}):
$\epsilon^2_{\,\mathrm{F}}=x^2_{\,\mathrm{F}}+1$. Then one obtains
the electron number density to be
$n_{e^-}=2\gamma/[(2\pi)^2\lambda^3_e]\, x_{\,\mathrm{F}}$ and the
matter density of the system of a one sort of nuclei $\rho_m=
\mu_{e^-} m_U n_{e^-}=(n_{e^-}/Z) m_A$, where $\mu_{e^-}=A/Z$ is the
molecular weight per electron [A(Z) is the mass (atomic) number of
the nucleus], $m_U$ is the atomic mass unit and $m_A$ is the mass of
the nucleus with the mass number $A$. For the lightest nuclei,
$\mu_{e^-}=2$, but, e.g., for $^{56}_{26}$\,Fe one obtains
$\mu_{e^-}=2.15$. The pressure of the degenerate electron gas is
then
\be P_{e^-}= \frac{\gamma m_e c^2}{(2\pi)^2\lambda^3_e}
\,\left[x_F\, \epsilon_{\,\mathrm{F}}-
ln(x_{\,\mathrm{F}}+\epsilon_F) \right]\,. \label{PNUT} \ee
With the choice  $\gamma$=$\epsilon^2_{\,\mathrm{F}}/2$ we stay at
the ground Landau level. Since $\epsilon^2_{\,\mathrm{F}}\gg$ 1 and
$x_{\,\mathrm{F}}\gg$ 1, the pressure of the degenerate electron gas
(\ref{PNUT}) can be written in the polytropic form:
%
\begin{equation}
P_{e^-}\,=\,K\, \rho_m^{\Gamma}\,, \qquad
\quad K\,= \frac{\pi^2\hbar^3}{(m_e\, m_U\,\mu_e)^2\,c\,\gamma}\,,
\qquad
\quad \Gamma\,=\,1+ \frac{1}{n} =\,2 \,,
\end{equation}
from which one obtains the value of the
polytropic index $n=1$. The input parameters of our study are
presented in Table \ref{tab:inpt}. Next we briefly describe the
calculation of the capture rate of the charge exchange reaction
(\ref{EMEPC}).
\begin{table}
\begin{tabular}{lcccc}
\hline
    \tablehead{1}{c}{b}{$\epsilon_{\,\mathrm{F}} $}
  & \tablehead{1}{c}{b}{${\rm n}_{e^-}/10^{33}[1/{\rm cm}^3$]}
  & \tablehead{1}{c}{b} {$\rho_{e^-}/10^6 [{\rm g/cm}^3]$}
  & \tablehead{1}{c}{b} {$\rho_m/10^{10} [{\rm g/cm}^3]$}
  & \tablehead{1}{c}{b} {$2\, \gamma$}  \\
\hline
 20        & 3.52 & 3.20 & 1.26 & 400   \\
 46        & 42.8 & 13.4 & 15.2 & 2116  \\
 90        & 321  & 293  & 114  & 8100  \\
 150       &1485  &1352  & 531  &22500  \\
 200        &3519  &3206  &1250  &40000  \\
\hline
\end{tabular}
\caption{The values of the Fermi energy
$\epsilon_{\,\mathrm{F}}$=E$_{\,\mathrm{F}}$/m$_e$c$^2$, used in the
present study. Further n$_{e^-}$ is the electron number density,
$\rho_{e^-}$ is the corresponding electron density, $\rho_m$ is the
matter density, calculated  for the nuclei $^{56}_{26}$\,Fe, and the
values of $\gamma$ are the smallest values allowing one to stay at
the ground Landau level.}
\label{tab:inpt}
\end{table}

\section{Reaction rate}

The $(e^-,e^+)$ conversion rate is, like the $0\nu\beta\beta$-decay
rate, proportional to the squared absolute value of the effective
mass of Majorana neutrinos $|\langle m_\nu\rangle|^2$. This quantity
is defined as $\langle m_\nu \rangle =  \sum_{i=1}^{3} U^2_{ei} m_i\,$, where $U$ is
the $3\times3$ Pontecorvo-Maki-Nakagawa-Sakata unitary mixing matrix
and $m_i$ ($i=1,2,3$) is the mass of the i-th light neutrino. The
$(e^-,e^+)$ conversion on nuclei is here considered  only for the
ground state to ground state transition, which is assumed to give
the dominant contribution. Both ground states of the initial
(${^{56}Fe}$) and the final (${^{56}Cr}$) nuclei have the spin and
parity $0^+$. The Coulomb interaction of electron and positron with
the nucleus is taken into account by the relativistic Fermi
functions $F(Z,E_{e^-})$ and $F(Z-2,E_{e^+})$ \cite{doi},
respectively. The leading order $(e^-,e^+)$ conversion matrix
element reads:
\begin{eqnarray}
  \label{S-matrix}
    \langle f \vert S^{(2)} \vert i \rangle &=&     2 \pi \delta(E_{e^+}-E_{e^-}
    + E_f - E_i) \langle f \vert T^{(2)} \vert i \rangle\,,
    \label{SM}\\
 \label{T-matrix}
\langle f \vert T^{(2)} \vert i \rangle &=&  \mathrm{i} ~\langle
m_\nu\rangle^* ~ \frac{1}{4 \pi} G^2_\beta \sqrt{F_0(Z,E_{e^-})}
\sqrt{F_0(Z-2,E_{e^+})}~ \overline{v}(P_{e^+}) (1 + \gamma_5)
u(P_{e^-})\times \frac{ g^2_{\mathrm{A}}}{R} {M}^{(e\beta^+)}\,.
\label{TM}
\end{eqnarray}
Here $G_\beta= G_{\,\mathrm{F}}\cos\theta_c$ and $E_i$ ($E_f$) is
the energy of the initial (final) nuclear ground state. The
conventional normalization factor of the nuclear matrix element
(NME) ${M}^{(e\beta^+)}$ involves the nuclear radius $R
=1.2~A^{1/3}~{\mathrm{fm}}$. For the weak axial coupling constant
$g_A$, we adopt the value $g_A=1.269$.

From Eq.\,(\ref{TM}), one obtains the following equation for the
reaction rate in the SMWDs:
\bea \Gamma^{(e\beta^+)} & =& m_e~
\frac{|\langle m_\nu\rangle|^2}{m^2_e} ~ \frac{1}{16 \pi^3}
{\left(\frac{G_{\beta} m^2_e}{~\sqrt{2}}\right)}^{{4}}
\frac{g^4_A}{(R^2 m_e^2)}~\left|{M}^{(e\beta^+)}\right|^2
\phi(\epsilon_{\,\mathrm{F}},\gamma)\,,  \label{RRPNF}
\eea
where the function $\phi(\epsilon_{\,\mathrm{F}},\gamma)$ is defined
as:
\bea
\phi(\epsilon_{\,\mathrm{F}},\gamma)&=&\frac{2\gamma}{(2\pi)^2\lambda^3_e
m^3_e}\,\int\limits_{Q+1}^{\epsilon_{\,\mathrm{F}}}\,
\left[\frac{(\epsilon_{e^-}-Q)^2-1}{\epsilon_{e^-}-1}\right]^{1/2}
\,(\epsilon_{e^-}-Q)\,\epsilon_{e^-}\,
F_0(Z,\epsilon_{e^-})~F_0(Z-2,\epsilon_{e^+}) d\epsilon_{e^-}\, ,
\label{ffi} \eea
with $\epsilon_{e^\pm}=E_{e^\pm}/m_e c^2$ and Q=$\Delta/m_e c^2$.\\

To calculate nuclear matrix element for the transition
(e$^-$,e$^+$) on $^{56}$Fe we use the Quasiparticle Random
Phase Approximation (QRPA)
\cite{src09} . For the A=56 system, the single-particle
model space consisted of $0-4\hbar\omega$ oscillator shells, both
for the protons and neutrons. The single particle energies are
obtained by using a  Coulomb--corrected Woods--Saxon potential. We
derive the two-body G-matrix elements from the Charge Dependent Bonn
one-boson exchange potential \cite{CDB} within the Brueckner theory.
For the quantitative analysis of the (e$^-$,e$^+$) capture rate we
will consider $|{M}^{(e\beta^{+})}| \approx 3\,.$

To estimate the energy production $\bar{\varepsilon}_r$ per
one event of the reaction (\ref{EMEPC}), we calculated the
two-photon positron-electron annihilation
probability per volume normalized to unity and integrated it over the energies of
electrons $\epsilon_f=E_f/m_e$ interacting with the positron in the
final state of reaction. From this and Eq.\,(\ref{RRPNF}), one can
obtain directly the released energy per 1 second as a contribution
to the luminosity. Here we made the calculations for the case of the
SMIWD with the mass M$_{\mbox{\tiny{WD}}}$=2\,M$_\odot$ and
 present them in Table \ref{tab:res}  as the ratio of the calculated change
in the luminosity $\Delta\, $L to the solar luminosity L$_\odot$.
\begin{table}
\begin{tabular}{lccc}
\hline
    \tablehead{1}{c}{b}{$\epsilon_{\,\mathrm{F}} $}
  & \tablehead{1}{c}{b}{$\left[Log(\Delta L/L_\odot)\right]_{0.4}$ }
  & \tablehead{1}{c}{b}{$\left[Log(\Delta L/L_\odot)\right]_{0.8}$}
  & \tablehead{1}{c}{b}{$\bar{\varepsilon}_r\, [{\rm MeV}]$} \\
\hline
 20        & -15.1 & -14.4   &  9.1   \\
 46        & -12.0 & -11.4   &  25.2  \\
 90        & -10.0 &  -9.4   &  49.1  \\
 150       &  -8.6 &  -8.0   &  79.7  \\
 200       &  -7.9 &  -7.3   & 104.4  \\
\hline
\end{tabular}
\caption{The values of the Fermi energy
$\epsilon_{\,\mathrm{F}}$=E$_{\,\mathrm{F}}$/m$_e$, used in the
present study. Further  $\Delta$L/L$_\odot$  is the ratio of the
change in the luminosity of the SMIWD due to the reaction
(\ref{EMEPC}) to the luminosity of the Sun, calculated for $|\langle
m_\nu\rangle|$=0.4 eV and 0.8 eV, and $\bar{\varepsilon}_r$ is the
 energy, released in single reaction.}
\label{tab:res}
\end{table}

\section{Cooling of white dwarfs}

To see how the energy, produced by the process (\ref{EMEPC}),
influences the cooling of SMIWDs, one should include it into
appropriate detailed microscopic model. Unfortunately, such detailed
models of cooling for this kind of white dwarfs have not yet been
elaborated.

In order to proceed, we first calculated  the luminosity of the iron
WDs within a simple cooling model, formulated by Mestel \cite{LM}.
We employed the electron pressure in the polytropic form with the
polytropic index n=3/2 and Kramers' opacity. Following the standard
calculation of the cooling rate \cite{LM,ST} one gets the relation
between the luminosity and the cooling time,
\be
\frac{L}{L_\odot}\,=\,1.3\,\times\,10^{-4}\,\left(\frac{M}{M_\odot}\right)\,
\left(\frac{Gyr}{\tau}\right)^{7/5}\,.\label{RLT} \ee
Then we calculated the luminosity of the SMIWDs in the same Mestel's
model. For the ground Landau level, the electron pressure is also of
the polytropic form, but with the polytropic index n=1; also for
this calculation we used Kramers' opacity with the result for the
luminosity,
\be
\frac{L}{L_\odot}\,=\,3.5\,\times\,10^{-2}\,\left(\frac{M}{M_\odot}\right)\,
\gamma^{\, 4/7}\,\left(\frac{Gyr}{\tau}\right)^{9/7}\,. \label{SMILT}
\ee
Numerical calculations show that both results differ significantly,
most probably because of the difference in the opacity, since the
luminosity of both types of white dwarfs are not likely to vary so
much. Thus, simple Mestel's model does not provide a reliable
estimate of the role of the considered reaction.

In order to estimate qualitatively possible effect of the reaction
(\ref{EMEPC}) on the cooling of the SMIWDs, we addressed the
asymptotic of the luminosity curves, obtained in Ref.\,\cite{PAB}
for the iron-core WDs. It is clear from Table \ref{tab:res} that the
effect of the double charge exchange reaction (\ref{EMEPC}) could
influence the cooling only at low luminosity. So extrapolating the
data, presented in Fig.\,17 \cite{PAB} for the curve, corresponding
to $M/M_\odot$=0.6 to smaller values of the luminosity, we got that
$Log(L/L_\odot)\,\approx\,$-5.0 (-7.54) is achieved after the
cooling time $\tau\,\approx\,$3.48 (3.90) Gyr\footnote{We obtained
similar results also extrapolating the data for the curve,
corresponding to $M/M_\odot$=0.8.}. One can see that the value
$Log(L/L_\odot)$=-7.54 is of the same size as are the values of
$Log(\Delta L/L_\odot)$, presented in the last line of Table
\ref{tab:res}. It means that the reaction (\ref{EMEPC}) could
effectively retard the cooling of the SMIWDs after low
enough luminosity evolves, unless are meanwhile all the iron nuclei  transformed
into chrome ones. However, before making more definite conclusions, the
theory of cooling of the SMIWDs should be elaborated.

\begin{theacknowledgments}
This work was supported  by the Votruba-Blokhintsev Program for
Theoretical Physics of the Committee for Cooperation of the Czech
Republic with JINR, Dubna. F. \v S. acknowledges the support by
the VEGA Grant agency
of the Slovak Republic under the contract No. 1/0876/12.
We thank L.~Althaus for providing us with
the data, presented in Fig.\,17 \cite{PAB} by the luminosity curves
for the pure iron-core DA WDs.
\end{theacknowledgments}




\bibliographystyle{aipproc}   

\bibliography{bibliography}

\IfFileExists{\jobname.bbl}{}
 {\typeout{}
  \typeout{******************************************}
  \typeout{** Please run "bibtex \jobname" to optain}
  \typeout{** the bibliography and then re-run LaTeX}
  \typeout{** twice to fix the references!}
  \typeout{******************************************}
  \typeout{}
 }

\end{document}